\documentclass[showpacs,twocolumn,prb]{revtex4}
\usepackage{graphicx}
\newcommand{\be}{\begin{equation}}
\newcommand{\ee}{\end{equation}}
\newcommand{\bea}{\begin{eqnarray}}
\newcommand{\eea}{\end{eqnarray}}
\newcommand{\bwt}{\begin{widetext}}
\newcommand{\ewt}{\end{widetext}}

\begin{document}
\title{Intraband Optical Spectral Weight in the presence of a van Hove singularity:
application to Bi$_2$Sr$_2$CaCu$_2$O$_{8+\delta}$}

\author{F. Marsiglio$^{1,2}$, F. Carbone$^1$, A. Kuzmenko$^1$, D. van der Marel$^1$}
\affiliation{ $^1$DPMC, Universit\'e de Gen\`{e}ve, 24 Quai
Ernest-Ansermet, CH-1211 Gen\`{e}ve 4, Switzerland, \\
$^2$Department of Physics, University of Alberta, Edmonton, Alberta,
Canada, T6G~2J1}

\begin{abstract}
The Kubo single band sum rule is used to determine the optical
spectral weight of a tight binding band with further than nearest
neighbour hopping. We find for a wide range of parameters and doping
concentrations that the change due to superconductivity at low
temperature can be either negative or positive. In contrast, the
kinetic energy change is always negative. We use an ARPES determined
tight binding parametrization of Bi$_2$Sr$_2$CaCu$_2$O$_{8+\delta}$
to investigate whether this can account for recent observations of a
positive change in the spectral weight due to the onset of
superconductivity. With this band structure we find that in the
relevant doping regime a straightforward BCS calculation of the
optical spectral weight cannot account for the experimental
observations.
\end{abstract}

\date{\today}
\maketitle

\section{introduction}

Recent optical experiments in several high $T_c$ cuprates at optimal
and low doping levels
\cite{molegraaf02,santander-syro03,kuzmenko05,carbone06a} have shown
an increase in the low frequency spectral weight when the system
goes superconducting. These observations are at odds with the
simplest expectation based on BCS theory
\cite{vandermarel03,benfatto05,marsiglio06}, where the kinetic
energy is expected to increase in the superconducting state;
however, they conform with the general notion of "kinetic
energy-driven" superconductivity \cite{hirsch92}. Since the optical
spectral weight is just the negative of the kinetic energy for a
single band with nearest neighbour hopping only, a {\em decrease} in
spectral weight is expected to occur below the superconducting
transition temperature. Several alterations to the standard BCS
picture have been proposed, the most minor of which involve an
alteration to the boson spectrum when the system goes
superconducting \cite{norman02,knigavko04,schachinger05,knigavko05}.

More recent measurements \cite{bontemps06,carbone06b} have shown a
continuous evolution with doping from "non-BCS-like" (low doping) to
"BCS-like" (high doping) behaviour; an understanding of this doping
dependence has been suggested in Ref.
\onlinecite{haule06,carbone06b} based on strong electron-electron
correlations and in Ref. \onlinecite{toschi05} based on the doping
dependence of a transition driven by pairing vs. phase coherence.

While all these proposals remain interesting possibilities to
explain the observations, the purpose of this paper is to revisit
the question of what "BCS-like" behaviour is, by taking into account
band structure details. It is important to do this, since our
intuition is based on the behaviour of the kinetic energy (which, as
we illustrate below) always increases in the superconducting state.
However, the optical spectral weight for a single band is given by
\cite{kubo57,maldague77}
\begin{equation}
W(T) = {2 \hbar^2 \over \pi e^2}\int_{0}^{+\infty }d\nu \mathop{\rm
Re} \left[ \sigma_{\rm xx} (\nu )\right] = \frac{2}{N}\sum_{k}
{\partial^2\epsilon_k \over \partial k_x^2} n_{k},%
\label{sumrule}
\end{equation}
whereas the negative of the band kinetic energy is given by a
somewhat different expression; in the simplest case it is given by

\begin{equation}
-<K> =  -\frac{2}{N}\sum_{k} \epsilon_k  n_{k},%
\label{kinetic}
\end{equation}
where $\epsilon_k$ is the tight-binding dispersion (that takes into
account already Hartree-Fock-type corrections) and $n_k$ is the
single spin momentum distribution function (we take the lattice
spacing to be unity). The sum over $k$ is over the first Brillouin
zone, and in the case with bilayer splitting (see below), includes a
summation over the two split bands. Note that this is {\em not} the
total kinetic energy of all the electrons, but just the kinetic
energy of the electrons in the given tight-binding band(s);
furthermore, {\em only} in the case of nearest neighbour hopping is
$W$ proportional to $-<K>$. In the presence of more complicated
interactions, the expectation value of the kinetic energy has more
complicated terms.

We first review the expectation for the kinetic energy, based on Eq.
(\ref{kinetic}), since this correspondence has been used to build
intuition concerning the optical spectral weight. First, what
happens when the system goes superconducting? The momentum
distribution function changes as discussed previously
\cite{marsiglio06}--- it goes from a Fermi-like distribution
function (in the absence of strong correlations) to a distribution
smeared by the presence of a superconducting order parameter. For an
order parameter with d-wave symmetry, the momentum distribution is
no longer a function of the band structure energy, $\epsilon_k$
alone. For example, for a BCS order parameter with simple nearest
neighbour pairing form, $\Delta_k = \Delta (\cos{k_x} -
\cos{k_y})/2$, then, as $k$ varies from $(0,0)$ to $(\pi,0)$, the
magnitude of the order parameter changes from zero to $\Delta$. On
the other hand, as $k$ varies along the diagonal (from the bottom of
the band to the top), the order parameter is zero (and constant). In
any event, even at zero temperature, BCS-like superconductivity {\em
raises} the kinetic energy of the electrons (see Fig. (1b) of Ref.
\onlinecite{marsiglio06}). This is as expected, since for
non-interacting electrons the normal state at zero temperature
corresponds to a state with the lowest possible kinetic energy.
Therefore, any modification to this state (for example, because of a
superconducting instability) can only increase the kinetic energy
expectation value.

\begin{figure}[tp]
\begin{center}
\includegraphics[height=5.4in,width=4.0in]{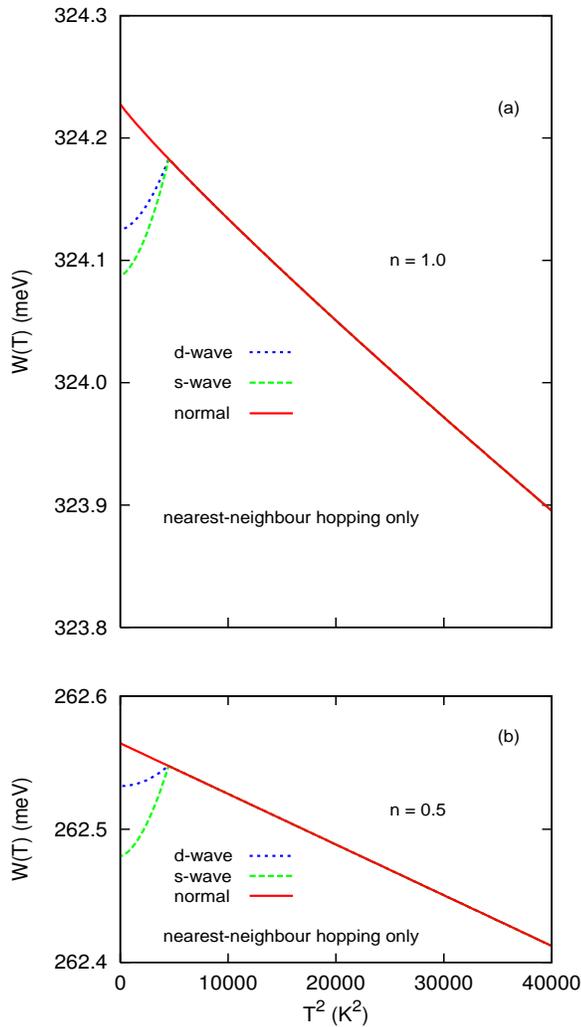}
\caption{Spectral Weight vs. $T^2$ for (a) half-filling, and (b) $n
= 0.5$. The normal state is given by the solid red curve, and the
superconducting state with d-wave (s-wave) symmetry by the
short-dashed blue (dashed green) curve. In both cases the normal
state result is almost linear in $T^2$, and the superconducting
state shows a decrease in the spectral weight (increase in the
kinetic energy) as expected. We used $t = 0.4$ eV, and BCS values
for order parameters, etc. with $T_c = 69$ K.}
\end{center}
\end{figure}

The question, partially answered in Refs.
\onlinecite{marsiglio06,carbone06b} is: does this behaviour remain
at all electron densities? Furthermore, with further than nearest
neighbour hopping, does the spectral weight (given by Eq.
(\ref{sumrule})) also follow the same trend as the negative of the
kinetic energy? Perhaps not surprisingly, we find that the spectral
weight does not qualitatively follow the kinetic energy near a van
Hove singularity. However, as will be discussed further below, we
find that for the band structure and doping regime thought to be
applicable in Bi$_2$Sr$_2$CaCu$_2$O$_{8+\delta}$ (BSCCO)
\cite{kordyuk03}, the spectral weight should decrease in the
superconducting state relative to the normal state. That is,
correlations, phase fluctuations, scattering rate collapse, or some
other scenario is required to understand the "anomalous" behaviour.
We will also address the temperature dependence in the normal state;
in some ways this is a more easily measured quantity than the change
below $T_c$.

In the next section we examine the optical spectral weight for a
model with nearest neighbour hopping only. This simple band
structure yields an optical spectral weight which is directly
proportional to the expectation value of the negative of the kinetic
energy. We examine the behaviour of the optical spectral weight as a
function of electron density. Note that we will use the symbol $n$
to denote electron density; for a single band this quantity will
span values from 0 to 2. It will be used when systematic
investigations of the spectral weight for a given band structure are
carried out. When comparing with experiments in the cuprates, we
will use the symbol $\delta$ to denote doping away from
half-filling, i.e. $n = 1$. Thus, $\delta = 1 - n$, and the regime
of experimental interest is roughly $0 < \delta < 0.25$. We use the
phrase "hole doping" to refer specifically to the value of $\delta$.

Following this section we introduce next-nearest neighbour hopping
into the band structure ($t-t^\prime$ model). This moves the van
Hove singularity away from half filling and also causes the spectral
weight to deviate from the kinetic energy; hence both will be
plotted in the ensuing plots. We find already in this simple
extension significant departures from the "standard BCS" description
based solely on the kinetic energy.

Finally, following Ref. \onlinecite{kordyuk03}, we also introduce a
next-next-nearest neighbour hopping and a bi-layer splitting term;
these are required for a quantitatively accurate description of the
ARPES (Angle-Resolved PhotoEmission Spectroscopy) results. We find
that these terms have significant effects on the optical sum rule.
First, the van Hove singularity is split into two singularities,
secondly the first of these occurs at a much lower hole doping level
than in the $t-t^\prime$ model.

As discussed in the summary, the end result is that (i) the change
in the optical spectral weight due to superconductivity can be
either positive or negative, depending on the band structure and
electron density, and (ii) if a parametrization of the band
structure is adopted from ARPES studies, then the optical spectral
weight {\em decreases} in the superconducting state. The observed
{\em increase} for optimal and underdoped samples then requires
additional ingredients. Some possibilities are briefly mentioned.

\section{nearest neighbour hopping only}

For nearest neighbour hopping only, the band structure is given by
\begin{equation}
\epsilon_k^{nn} = -2t(\cos{k_x} + \cos{k_y})%
\label{nn}%
\end{equation}
and we have that $2W = -<K>$ in two dimensions. In Fig. 1 we show
plots of the spectral weight vs. $T^2$ for two representative
electron densities, $n = 1$ and $n = 0.5$. The first places the
Fermi level right on the van Hove singularity, while the second is
well removed from all van Hove singularities. These are computed
through the usual procedure: first, even in the normal state, the
chemical potential must be determined at each temperature to ensure
that the electron density remains constant as a function of
temperature. This is the common procedure, though it is true that in
complicated systems for which one is using some "low energy"
tightbinding Hamiltonian to describe the excitations that it is not
immediately obvious that the electron number density should remain
fixed as a function of temperature; nonetheless, we adopt this
procedure here. In Eq. (\ref{sumrule}) the chemical potential enters
the momentum distribution function, which, in the normal state, is
replaced by the Fermi-Dirac distribution function, $n_k \rightarrow
f(\epsilon_k - \mu)$. In the superconducting state, we simply adopt
a model temperature dependence for the order parameter, following
Ref. \onlinecite{marsiglio06}. This has been tested for both s-wave
and d-wave symmetries by comparing to fully self-consistent
solutions to BCS equations with separable potentials
\cite{marsiglio06}. One still has to determine the chemical
potential self-consistently for each temperature, which is done by
solving the number equation in the superconducting state for a fixed
chemical potential and order parameter, and iterating until the
desired number density is achieved. The value of the zero
temperature order parameter is fixed by the weak coupling BCS
values, $2\Delta_0 = \eta k_BT_c$ where $\eta = 4.2 (3.5)$ for the
d-wave (s-wave) case. Further details are provided in Ref.
\onlinecite{marsiglio06}. For the electron densities studied in the
first part of this paper, we simply take $T_c = 69$ K for all
electron densities. In section IV we will adopt $T_c$ values as
observed from experiment.

Both plots in Fig. 1 show somewhat linear behaviour with $T^2$,
though in Fig. 1(a) there is some noticeable upward curvature due to
the van Hove singularity which is present at the Fermi level for
this electron density. The decrease in spectral weight at the
transition is more pronounced for s-wave symmetry (dashed green
curves) than for d-wave symmetry (dotted blue curves). The normal
state results show a decreasing value with increasing temperature,
indicative of an increasing kinetic energy. This is the 'textbook'
example of the temperature dependence of the spectral weight through
a superconducting transition
\cite{vandermarel03,benfatto05,marsiglio06}.

\begin{figure}[tp]
\begin{center}
\includegraphics[height=3.4in,width=3.4in]{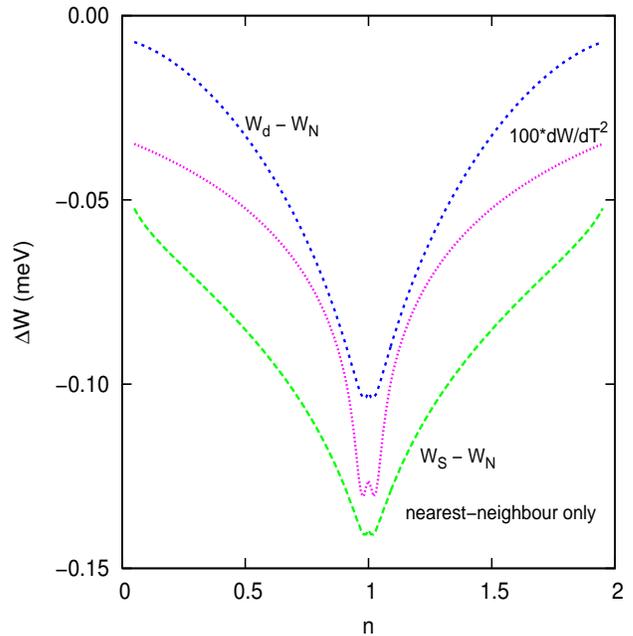}
\caption{The difference ($W_d - W_N$) in the spectral weight between
the superconducting state with d-wave symmetry and the normal state
at zero temperature vs. doping (dotted blue curve). The dashed green
curve shows the same quantity for s-wave symmetry, and the pink
points indicate the slope (with respect to $T^2$) of the spectral
weight near $T_c$. All three quantities are always negative, and
show an enhancement near half-filling due to the van Hove
singularity. In fact, the pink points are almost a perfect inverted
image of the density of states (see the minus sign in Eq.
(\protect\ref{kin_corr})), except for the small density regime near
half-filling, where the van Hove singularity makes the Sommerfeld
expansion invalid.}
\end{center}
\end{figure}

In Fig. 2 we examine both the spectral weight difference ($W_s -
W_N$ and $W_d - W_N$ for s- and d-wave symmetry, respectively ---
'N' here stands for 'normal') at zero temperature, and the slope of
$W(T)$ with respect to $T^2$ at $T_c$, vs. electron density, n.
These plots make evident several important points. First, the van
Hove singularity clearly plays a role; it enhances the overall
magnitude of the effect, whether we examine the difference between
the superconducting and normal state at zero temperature, or the
slope at $T_c$. In fact the latter tracks the former, indicating
that both are related to one another. One can understand this
qualitatively by the observation that in both cases (warming up or
going superconducting) the momentum distribution function broadens,
though for different reasons \cite{marsiglio06}. The most important
point to learn from this plot is that the difference is always
negative, indicating that, for nearest neighbour hopping only,  the
opening of a gap does indeed increase the kinetic energy and
decrease the spectral weight in a superconductor.

\section{next-nearest neighbour hopping}

When next-nearest neighbour hopping is included in the band
structure, one obtains the so-called $t-t^\prime$ model. This model
has band structure
\begin{equation}
\epsilon_k^{nnn} = -2t(\cos{k_x} + \cos{k_y}) + 4t^\prime \cos{k_x}
\cos{k_y},%
\label{ttprime}
\end{equation}
and goes a long way towards understanding the Fermi surface for
Bi$_2$Sr$_2$CaCu$_2$O$_{8+\delta}$ (BSCCO), as determined by ARPES
\cite{norman95,kordyuk03}, at least for the doping levels studied.
On the theoretical side, the presence of $t^\prime$ shifts the van
Hove singularity to an energy given by $\mu = -4t^\prime$. For the
sake of this study one can study all electron densities; however,
one must bare in mind that most experiments on BSCCO are at doping
levels such that the van Hove singularity is not crossed, i.e. the
Fermi surfaces are always hole-like. We will also study (see next
section) a band structure more pertinent to BSCCO
\cite{norman95,kordyuk03}, which uses a next-next-nearest neighbour
hopping amplitude in addition:
\begin{equation}
\epsilon_{k \pm} = \epsilon_k^{nnn} - 2t^{\prime \prime}(\cos{2k_x} + \cos{2k_y})
\pm t_\perp(\cos{k_x} - \cos{k_y})^2/4,%
\label{ttprimettwoprime}
\end{equation}
which we will refer to as the $t-t^\prime-t^{\prime \prime}$ model.
Note that we allow for a bilayer splitting term as well, following
Kordyuk et al. \cite{kordyuk03}. However, they actually adjust
hopping parameters for each doping, while we simply adopt the ones
used for their overdoped sample: $t=0.40$ eV, $t^\prime = 0.090$ eV,
$t^{\prime \prime} = 0.045$ eV, and $t_\perp = 0.082$ eV.
Illustrative plots of the band structures are shown in Fig. 3.

\begin{figure}[tp]
\begin{center}
\includegraphics[height=4.9in,width=3.5in]{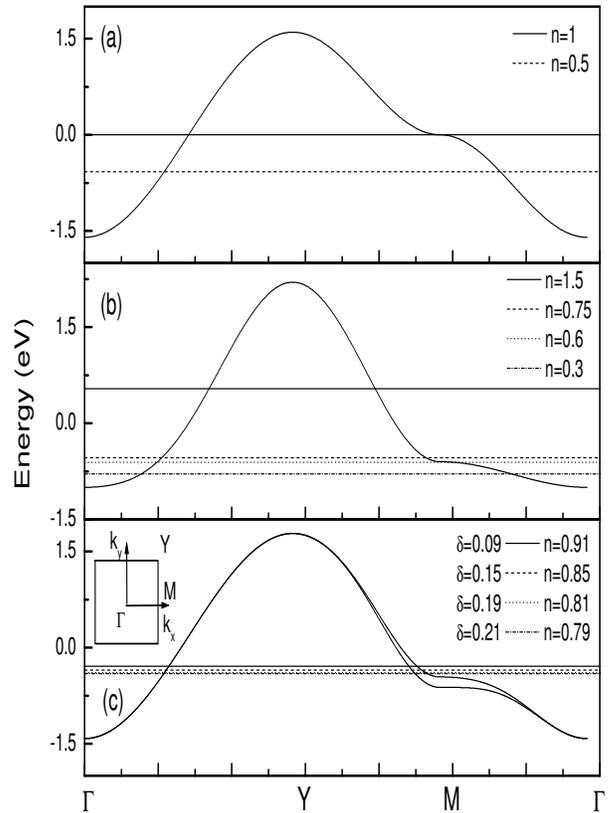}
\caption{Illustrative plots of the band structure for (a) nearest
neighbour hopping only, (b) the $t-t^\prime$ model, and (c) the
Kordyuk et al. \protect\cite{kordyuk03} parametrization of the band
structure with bilayer splitting. The van Hove singularities occur
where the band dispersion flattens.}
\end{center}
\end{figure}

Returning now to the $t-t^\prime$ model, the van Hove singularity
occurs at an electron density $n = 0.60$, i.e. a hole doping (away
from half-filling) of $\delta = 1 - n = 0.4$. As mentioned above,
this high level of doping is never realized in samples of BSCCO
\cite{ino02}. In any event, we are interested in the more generic
behaviour of the spectral weight, given a reasonably representative
band structure for the cuprates.

\begin{figure}[tp]
\begin{center}
\includegraphics[height=7.4in,width=4.1in]{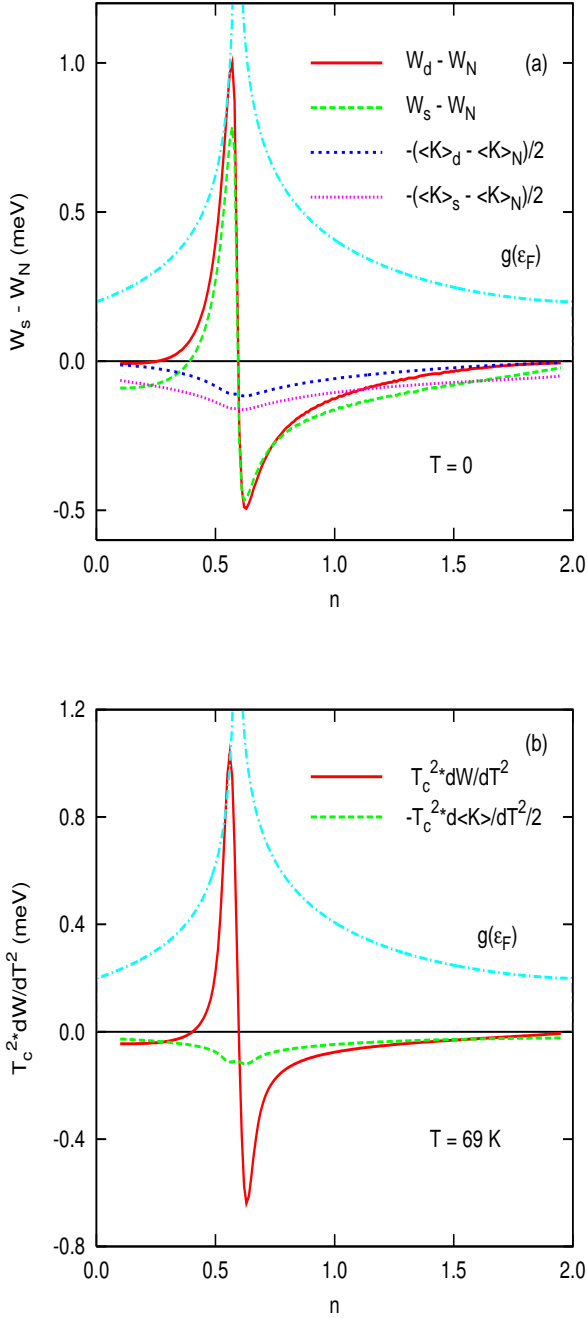}
\caption{(a) The difference ($W_d - W_N$) in the spectral weight
between the superconducting state with d-wave symmetry and the
normal state at zero temperature vs. doping (solid red curve), for
the $t-t^\prime$ band structure, with $t=0.4$ eV and $t^\prime =
0.09$ eV. The dashed green curve shows the same quantity for s-wave
symmetry. Both exhibit positive values to the left of the van Hove
singularity (the density of states at the Fermi level is indicated,
as a function of doping, by the dot-dashed cyan curve). The negative
of the kinetic energy for d-wave (blue short-dashed curve) and for
s-wave (dotted pink curve) behaves as expected, always negative, and
peaks (in absolute value) at the van Hove singularity. (b) The
normal state slope (taken at $T_c = 69$ K) of the spectral weight
vs. doping (solid red curve). The dashed green curve shows the same
quantity for the negative of the kinetic energy. These behave in
very similar fashion to the differences (taken at zero temperature)
shown in (a).}
\end{center}
\end{figure}

In Fig. 4 we show a summary of the doping dependence of the various
quantities with the $t-t^\prime$ band structure. In both Fig. 4a and
4b we have plotted the density of states at the Fermi level as a
function of doping (this is possible for a doping-independent band)
to illustrate where the van Hove singularity is. The remarkable
feature in Fig. 4a, for electron densities below (i.e. hole doping
away from half filling "above") the van Hove singularity, is that
the spectral weight change in the superconducting state is positive
! Similarly in Fig. 4b the actual slope of the spectral weight above
$T_c$ is positive. Note that our intuition about the kinetic energy
change remains correct; it is indeed negative, for all electron
densities, for both s-wave and d-wave symmetries. Moreover, the
slope is also everywhere negative, which establishes a definite
correlation between the slope above $T_c$ and the change at $T=0$.
Note that in Ref. \onlinecite{marsiglio06} (see Fig. 4 of that
reference) the doping parameters were such that the optical sum rule
and the negative of the kinetic energy were qualitatively (and even
quantitatively) similar \cite{remark1}. Here, in the vicinity of the
van Hove singularity we encounter a regime where these two
properties are markedly different from one another.

How do we understand this strange dependence near the van Hove
singularity ? It is easiest to focus on the temperature dependence
in the normal state. One should first review the simple argument for
why the slope for the negative of the kinetic energy is expected to
{\em always} be negative, no matter what the band structure. This
quantity is defined as
\begin{equation}
-<K> = - \int_{-\infty}^{+\infty} d \epsilon \ \epsilon \
g(\epsilon) f(\epsilon - \mu) \label{kinetic_2}
\end{equation}
where $g(\epsilon) \equiv {1 \over N} \sum_k \delta(\epsilon -
\epsilon_k)$ is the single spin, single particle density of states
for electrons with band dispersion $\epsilon_k$. One can perform a
Sommerfeld expansion, which yields two terms, one of which is
eliminated by performing the analogous expansion for the electron
density, with the requirement that the density remains constant as a
function of temperature \cite{ashcroft76}. The result is
\cite{ashcroft76}
\begin{equation}
-<K> = -<K>_{T=0} - {\pi^2 \over 3} (k_BT)^2 g(\epsilon_F),
\label{kin_corr}
\end{equation}
where, it is now clear that, regardless of the Fermi level, the
temperature correction is always negative. This means that the
kinetic energy ($<K>$) always increases as the temperature
increases, a statement which is physically obvious.

\begin{figure}[tp]
\begin{center}
\includegraphics[height=3.5in,width=3.5in]{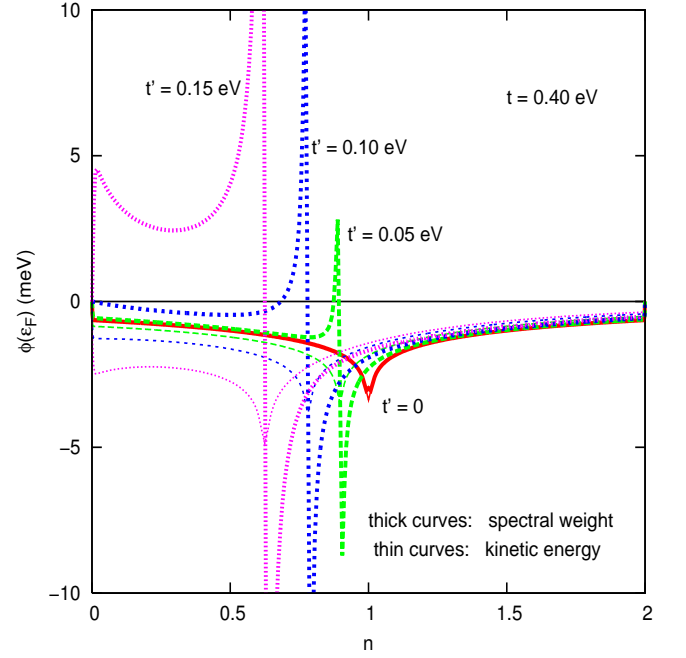}
\caption{The finite temperature correction to the optical spectral
weight (thick curves), for various values of $t^\prime$, as
indicated. Also plotted is the analogous quantity for the kinetic
energy (which, by Eq. (\protect\ref{kin_corr}) is just
$-8tg(\epsilon_F)$). Note that the latter is always negative, while
the spectral weight becomes positive for a finite electron
concentration for electron (hole) densities below (above) the van
Hove singularity. The magnitude of the anomalous region increases
with increasing $t^\prime$.}
\end{center}
\end{figure}

With the spectral weight,
\begin{equation}
W \equiv {2 \over N} \sum_k {\partial^2 \epsilon_k \over \partial
k_x^2} f(\epsilon_k - \mu), \label{spectral_weight}
\end{equation}
there is no simple cancelation, as occurs in the kinetic energy. If
one defines the quantity,
\begin{equation}
g_{xx}(\epsilon) \equiv {1 \over N} \sum_k {\partial^2 \epsilon_k
\over \partial k_x^2} \delta(\epsilon - \epsilon_k), \label{gxx}
\end{equation}
then the Sommerfeld expansion can be applied to $W(T)$ as was done
for the kinetic energy. The result is
\begin{equation}
W(T) = W(0) + {\pi^2 \over 3} {(k_BT)^2 \over 8t}
\phi(\epsilon_F),%
\label{spec_corr}%
\end{equation}
where
\begin{equation}
\phi(\epsilon_F) = 8t \bigg\{ {g_{xx}^\prime(\epsilon_F)
g(\epsilon_F) - g^\prime(\epsilon_F)g_{xx}(\epsilon_F) \over g(\epsilon_F)} \biggr\}.%
\label{spec_corr2}
\end{equation}
Using $g_{xx}(\epsilon_F) = -\epsilon_F g(\epsilon_F)$, which is
correct for nearest neighbour hopping only, one finds
$\phi(\epsilon_F) = -8tg(\epsilon_F)$ and one recovers the previous
result, given by Eq. (\ref{kin_corr}) for the optical spectral
weight as well as for the kinetic energy. Otherwise, the sign of the
correction is dependent on the quantity within the braces. This
quantity involves a difference between two quantities, both of which
contain singularities. This quantity is plotted in Fig. 5 for the
$t-t^\prime$ model for various values of $t^\prime$, along with the
corresponding quantity for the kinetic energy, which, by Eq.
(\ref{kin_corr}), is $-8tg(\epsilon_F)$. It is evident that as
$t^\prime$ increases, a larger anomaly over a wider range of
electron densities is expected. For more complicated band
structures, an assortment of Fermi energies (i.e. electron
concentrations) will exist for which the slope (and the change in
the superconducting state) is "anomalous", i.e. positive.

\begin{figure}[tp]
\begin{center}
\includegraphics[height=6.3in,width=3.4in]{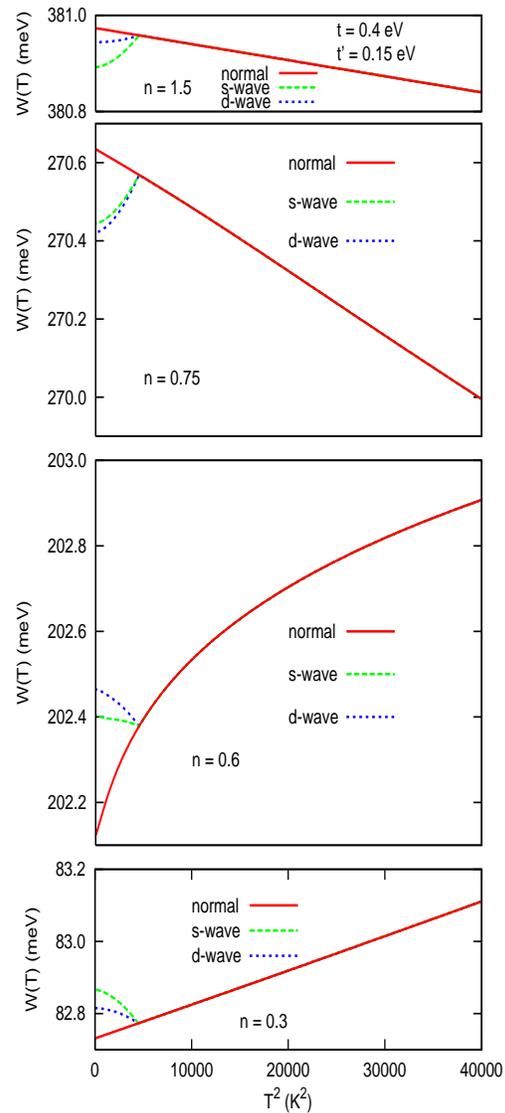}
\caption{Spectral weight for four representative densities for the
$t-t^\prime$ model (see Fig. 5). The progression from top to bottom
is from high electron concentration to low electron concentration.
Note the qualitative change from negative slope (lower
superconducting vs. normal state difference) to positive slope
(higher superconducting vs. normal state difference) as the electron
concentration changes such that the chemical potential sweeps
through the van Hove singularity.}
\end{center}
\end{figure}

In Fig. 6 we show some specific examples of the temperature
dependence of the optical spectral weight for a variety of electron
concentrations. These results clearly show the progression from the
'standard' result shown for $n=1.5$ (top frame) to the highly
anomalous result shown for $n=0.6$ (third frame). Note the variation
of the vertical scale: there is the obvious progression of lower
spectral weight with decreasing electron density. However, the
magnitude of both the slope and the change due to superconductivity
also experiences a large increase for the two electron densities
near the van Hove singularity ($n=0.75$ and $n=0.6$). For example,
the absolute value of the slope in the second frame ($n=0.75$ is
approximately a factor of 5 higher than that in the first frame
($n=1.5$). Similarly at $n=0.6$ the slope is much higher than at
$n=0.3$, particularly at low temperatures. The reason is evident
from Fig. 5: this enhancement occurs in the vicinity of the van Hove
singularity. Note that the results of Fig. 6 are displayed for
$t^\prime = 0.15$ eV. In this case, Fig. 5 shows that there is an
extended range of electron densities for which a positive slope (and
positive change below $T_c$) occurs. Using $t^\prime = 0.10$ eV, for
example, would result in a very narrow range of electron densities
for which the optical spectral weight has behaviour opposite to that
of the negative of the kinetic energy (see Fig. 5, blue dashed
curves).

In summary, in the case of the simple next-nearest neighbour model
considered, the anomalous region always occurs at electron
concentrations {\em below} the van Hove singularity. In many of the
high temperature superconducting cuprates, this electron
concentration is not experimentally achieved. Certainly in the
experiments that report a positive ("anomalous") optical spectral
weight change in BSCCO, this regime is believed to be irrelevant.
Hence, while this investigation of the behaviour of the optical
spectral weight using the $t-t^\prime$ model certainly plays havoc
with our intuition (which motivated the experiments in the first
place), it apparently does not provide an explanation of the
experimental results \cite{remark2}. This is further explored in the
next section.

\section{bi-layer splitting}

\begin{figure}[tp]
\begin{center}
\includegraphics[height=6.3in,width=3.4in]{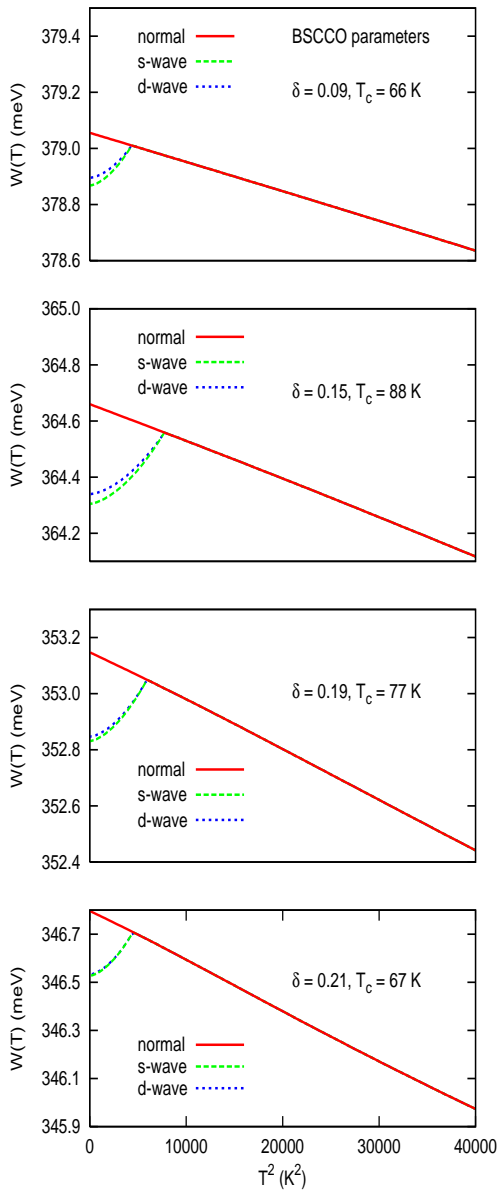}
\caption{Spectral weight for the four hole doping concentrations
considered in the experiments \protect\cite{carbone06b}. These
calculations use the ARPES-determined band structure parameters from
Ref. \protect\cite{kordyuk03}, including the bilayer splitting.
Slopes are always negative, and the change at $T_c$ is also always
negative. Note that in this case the s-wave order parameter gives
roughly the same value as the d-wave parameter.}
\end{center}
\end{figure}

It is evident that the characteristics of the optical spectral
weight will be very dependent on the band structure and the doping
level. A complete investigation of parameter space, with an
accompanying catalogue of optical spectral weight behaviour would
undoubtedly reveal a wide assortment of results. Perhaps a more
useful procedure at this point would be to adopt the band structure
proposed for a particular compound (BSCCO), as prescribed by some
other experiment (ARPES), and determine the behaviour of the optical
spectral weight in this case. As discussed in the previous section,
we will adopt the parameters used in Ref. (\onlinecite{kordyuk03})
to fit their ARPES results; these include a bilayer splitting term,
which, as will now be discussed, can result in even more unusual
doping dependence. Their determined structure was given in the
previous section by Eq. (\ref{ttprimettwoprime}) with parameters
listed below this equation. They also used $t_\perp = 0.082$ eV; we
will adopt these parameters for all electron concentrations (i.e.
hole doping) for simplicity.

\begin{figure}[tp]
\begin{center}
\includegraphics[height=3.4in,width=3.4in]{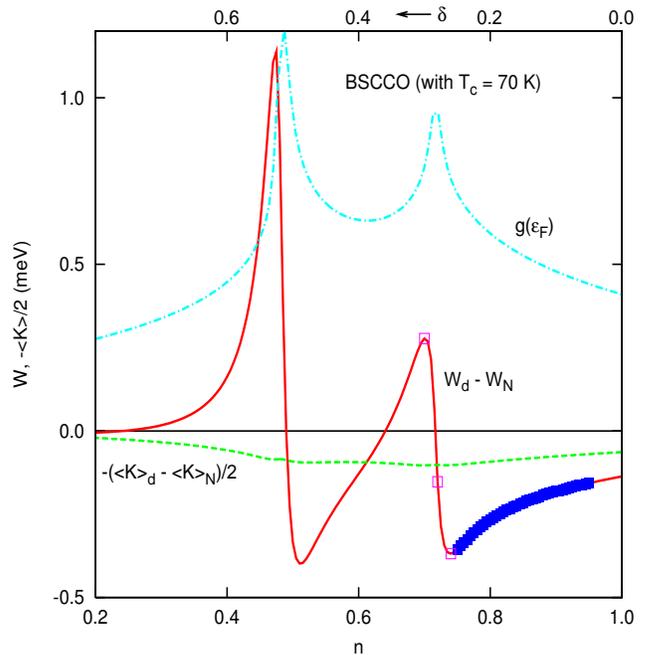}
\caption{Spectral weight difference (solid red curve) as a function
of electron concentration for the band structure determined by ARPES
\protect\cite{kordyuk03}. For simplicity we put $T_c = 70$ K for all
electron densities. The normal state slope shows very similar
behaviour. Also plotted is the negative of the change in the kinetic
energy (dashed green curve), and the density of states at the Fermi
level (dot-dashed cyan curve), for reference. Note that 'anomalies'
in the spectral weight occur at the two van Hove singularities,
while the change in kinetic energy is simply gently modulated by the
density of states. Also shown with solid blue symbols is the hole
doping regime of relevance to the BSCCO experiments in Ref.
\protect\onlinecite{carbone06b}, and considered in the previous
figure. For clarity only d-wave results are shown. The three open
squares show the doping concentrations considered in the next
figure, of possible relevance for overdoped samples with chemical
potential that crosses the van Hove singularity
\protect\cite{kaminski06}.}
\end{center}
\end{figure}

Fig. 7 shows the optical spectral weight as a function of
temperature for the four doping levels considered in the experiments
\cite{carbone06b,remark3}. It is clear that with the band structure
determined from ARPES \cite{kordyuk03}, the optical spectral weight
cannot be explained by the simple BCS model adopted here.

However, very recently, very overdoped BSCCO samples have been
produced \cite{kaminski06}; ARPES measurements \cite{kaminski06}
have determined that the chemical potential moves through the first
van Hove singularity. Thus, it is instructive to examine the
consequences of this BCS model for higher hole doping concentrations
(lower electron densities). In Fig. 8 the doping dependence of the
optical spectral weight slope is shown as a function of electron
concentration, $n$ for the hole doped region (with respect to
half-filling). This figure uses the band structure parameters from
Kordyuk et al. \cite{kordyuk03}, so the crossing of the first van
Hove singularity occurs at a higher hole doping (about $\delta
\approx 0.28$) than reported in Ref. \onlinecite{kaminski06}.
Leaving this detail aside for the moment, it is clear from Fig. 8
that a dramatic change is expected in the optical spectral weight
anomaly for this doping. In Fig. 9 we show the temperature
dependence for some hole doping concentrations near this region, as
indicated by the open squares in Fig. 8. Both the normal state and
superconducting state behaviours show a transition as one crosses
the van Hove singularity. It would be most interesting to measure
the optical spectral weight in this doping regime. Such measurements
would provide a good indication of whether or not the van Hove
singularity is relevant in these materials.

\begin{figure}[tp]
\begin{center}
\includegraphics[height=6.3in,width=3.4in]{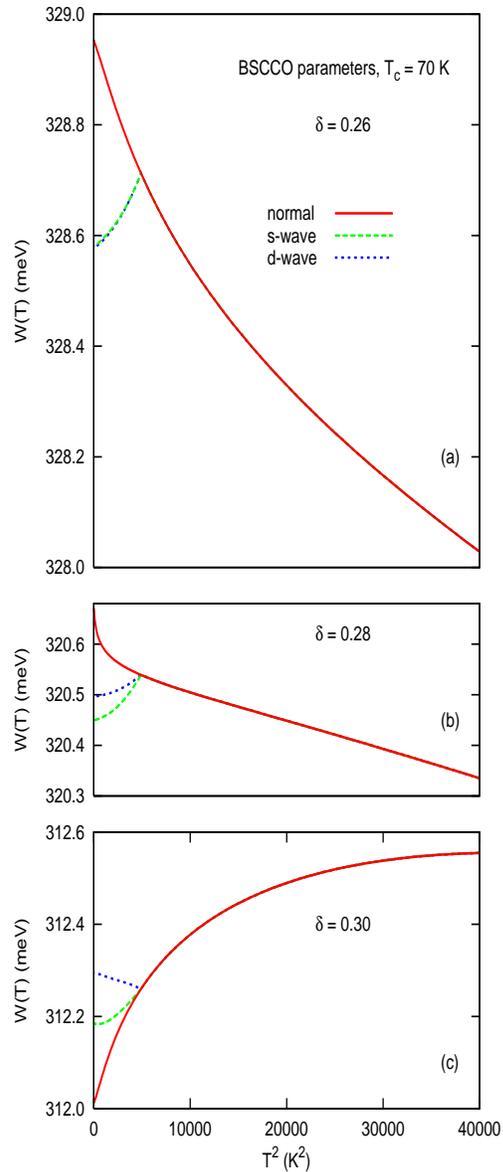}
\caption{Spectral weight for three hole doping concentrations that
move the chemical potential across the first van Hove singularity
(see Fig. 8). Note the dramatic change that occurs as one dopes
across the van Hove singularity (see text).}
\end{center}
\end{figure}

Returning to the samples whose optical properties have already been
measured, Figures 7 and 8 make clear that, taking the
ARPES-determined band structure seriously, then the observed doping
dependence of the optical spectral weight change below $T_c$
\cite{bontemps06,carbone06a,carbone06b} {\em cannot} be explained by
BCS theory. In this respect the observations may be indicative of
physics beyond BCS theory. This remains true even if the hole doping
is shifted by a small amount, that is, if the measured samples
actually have hole concentrations that are larger than $0.09 <
\delta < 0.21$; then the spectral weight anomaly calculated here has
almost the opposite doping dependence compared with experiment
\cite{carbone06b}.

On the other hand the band structure is a parametrization, and is
reliant on a very surface sensitive probe. As an example of what we
consider a remote possibility, Fig. 8 indicates that if the doping
levels for BSCCO are {\em not} as indicated, but rather lie in the
regime {\em between} the two van Hove singularities (i.e.
approximately between $n=0.5$ and $n=0.7$), then the results will be
very different. Then 'underdoped' would have a positive change below
$T_c$ (near $n \approx 0.7$) while 'overdoped' would exhibit a
negative change (compatible with experiment, and with expectations
based on the negative of the kinetic energy) (near $n \approx 0.6$).
The corresponding slope above $T_c$ would, however, be {\em
inconsistent} with experiment (not shown), but the slope is a purely
normal state property, and, like all other normal state properties,
undoubtedly requires electron correlations for a proper
understanding. We view this possibility as remote, however, since
many studies have established this band structure and the low doping
scenario.

A more reasonable explanation comes from including effects
considered in Ref. \onlinecite{marsiglio06}, along with the Kordyuk
et al. band structure. The proposal there was to include the effect
first observed in microwave measurements \cite{nuss91,kuroda90},
that the inelastic scattering rate decreases significantly below
$T_c$. This has the effect of shifting the blue points in Fig. 8,
for example, upwards. Thus, the anomaly below $T_c$ would agree with
experiment \cite{bontemps06,carbone06b}, including a crossover from
positive change for underdoped samples to a negative change for
overdoped samples. Furthermore, the slope above $T_c$ would always
be negative, as observed. Finally, the magnitude of the slope would
increase as the anomaly becomes more BCS-like, also in agreement
with experiment. This was also found in the DMFT calculations
\cite{carbone06b}, and further theoretical work and experiments
would be required, however, to disentangle band structure effects
from strong correlation effects. For example, the phenomenology of a
collapsed scattering rate is generally indicative of an electronic
scattering mechanism being responsible for superconductivity, so the
DMFT calculations may be reconstructing the same phenomenology; it
would be interesting to examine the interplay between strong
correlation an band structure effects in these calculations. We
should also remark that very early on Norman and P\'epin
\cite{norman02} found that using a frequency dependent scattering
rate extracted from optical data gave rise to a crossover as a
function of doping very similar to what has been observed in
experiment. Further work is required to distinguish between these
various scenarios.

\section{summary}

The primary result of this paper is the revelation that the single
band optical spectral weight may behave very differently from the
kinetic energy, both in the normal state, and in the superconducting
state. This occurs when one uses a band structure more complicated
than one involving nearest neighbour (nn) hopping only, since, with
nn hopping only, the two are identical. Thus, one cannot use the
phrase, 'BCS-like' behaviour for the optical spectral weight, but
one can continue to use that phrase for the kinetic energy.

This means that a wide variety of dependency on doping is in
principle possible, due to band structure effects alone. This has a
significant impact on the interpretation of experimental results, as
doping dependence due to correlation effects, for instance, would
have to be separated out either experimentally or theoretically.

On the other hand, if one accepts the band structure for, say,
BSCCO, as determined by ARPES, then the spectral weight observations
\cite{molegraaf02,santander-syro03,kuzmenko05,carbone06a,bontemps06,carbone06b}
remain anomalous, i.e. cannot be explained by BCS theory alone. We
have advanced a couple of possibilities, and many others have been
proposed in the literature: doping levels may be shifted slightly
compared to what we think they are, in which case strong
correlations well beyond BCS theory are required to explain the
observed trend with doping. A much stronger shift in doping actually
reproduces the observed trends in the superconducting anomaly, but
not in the normal state behaviour. Finally, including a scattering
rate collapse below $T_c$ also qualitatively accounts for the data.

\begin{acknowledgments}

FM greatly appreciates the hospitality of the Department of
Condensed Matter Physics at the University of Geneva, where this
research was performed. This work was supported in part by the
Natural Sciences and Engineering Research Council of Canada (NSERC),
by ICORE (Alberta), by the Canadian Institute for Advanced Research
(CIAR), and by the University of Geneva.
\end{acknowledgments}

\bibliographystyle{prb}

\begin{thebibliography}{1}


\bibitem{molegraaf02}  H.J.A.~Molegraaf, C.~Presura, D.~van~der~Marel,
P.H.~Kes, and M.~Li, Science {\bf 295}, 2239 (2002).

\bibitem{santander-syro03}
A.F.~Santander-Syro, R.P.S.M.~Lobo, N.~Bontemps, Z.~Konstantinovic,
Z.Z.~Li, H.~Raffy, Europhys. Lett. {\bf 62}, 568 (2003);
A.F.~Santander-Syro, R.P.S.M.~Lobo, N.~Bontemps, W.~Lopera,
D.~Girata, Z.~Konstantinovic, Z.Z.~Li, H.~Raffy, Phys.~Rev.~B {\bf
70}, 134504 (2004).

\bibitem{kuzmenko05} A.B.~Kuzmenko, H.J.A.~Molegraaf, F.~Carbone,
D.~van~der~Marel, Phys. Rev. B{\bf 72}, 144503 (2005).

\bibitem{carbone06a} F. Carbone, A.B.~Kuzmenko, H.J.A.~Molegraaf, E.
van Heumen, E. Giannini, and D.~van~der~Marel, submitted to Phys.
Rev. B cond-mat/0603737.

\bibitem{vandermarel03} D. van der Marel,  H.J.A. Molegraaf, C. Presura,
and I. Santoso, in {\it Concepts in Electron Correlations}, edited
by A. Hewson and V. Zlatic (Kluwer, 2003). See also
cond-mat/0302169.

\bibitem{benfatto05}
L. Benfatto, S. G. Sharapov, N. Andrenacci, and H. Beck, Phys. Rev.
B {\bf 71}, 104511 (2005).

\bibitem{marsiglio06} F. Marsiglio, Phys. Rev. B{\bf 73}, 064507
(2006).

\bibitem{hirsch92} J.E. Hirsch, Physica C 199, 305 (1992); Physica C 201, 347
(1992).


\bibitem{norman02}
M.R. Norman and C. P\'epin, Phys. Rev. B {\bf 66}, 100506 (2002).

\bibitem{knigavko04} A. Knigavko, J. P. Carbotte, and F.
Marsiglio, Phys. Rev. B {\bf 70}, 224501 (2004); Europhysics Letts.
{\bf 71}, 776 (2005).

\bibitem{schachinger05} E. Schachinger and J. P. Carbotte, Phys. Rev. B
{\bf 72}, 014535 (2005).

\bibitem{knigavko05} A. Knigavko and J.P. Carbotte, Phys. Rev. B {\bf 72},
  035125 (2005); cond-mat/0602681, to be published in Physical Review B.

\bibitem{bontemps06} N. Bontemps, R.P.S.M. Lobo, A.F.
Santander-Syro, and A. Zimmers, cond-mat/0603024.

\bibitem{carbone06b} F. Carbone, A.B.~Kuzmenko, H.J.A.~Molegraaf, E.
van Heumen, V. Lukovac, F. Marsiglio, D.~van~der~Marel, K. Haule, G.
Kotliar, H. Berger, S. Courjault, P.H. Kes, and M. Li, submitted to
Phys. Rev. B, cond-mat/0605209.

\bibitem{haule06} K. Haule and G. Kotliar, cond-mat/0601478.

\bibitem{toschi05}
A. Toschi, M. Capone, M. Ortolani, P. Calvani, S. Lupi, and C.
Castellani Phys. Rev. Lett. {\bf 95}, 097002 (2005); A. Toschi, M.
Capone, and C. Castellani, Phys. Rev. B {\bf 72}, 235118 (2005).

\bibitem{kubo57} R. Kubo, J. Phys. Soc. Japan {\bf 12}, 570 (1957).

\bibitem{maldague77} P.F. Maldague, Phys. Rev. B{\bf 16}, 2437
(1977).

\bibitem{norman95} M.R. Norman, M. Randeria, H. Ding, and J.C.
Campuzano, Phys. Rev. B{\bf 52}, 615 (1995).

\bibitem{kordyuk03} A.A. Kordyuk, S.V. Borisenko, M. Knupfer, and J. Fink,
Phys. Rev. B{\bf 67}, 064504 (2003).

\bibitem{ino02} Note that in the case of LSCO, doping can move the
chemical potential through the van Hove singularity; see A. Ino, C.
Kim, M. Nakamura, T. Yoshida, T. Mizokawa, A. Fujimori, Z.-X. Shen,
T. Kakeshita, H. Eisaki, and S. Uchida, Phys. Rev. B {\bf 65},
094504 (2002).

\bibitem{remark1} Note that the labels of Fig. 3 of Ref.
\onlinecite{marsiglio06} were inadvertently interchanged. Thus,
$\delta T_D$ and $\delta K_D$ should be switched, and similarly for
$\delta T_S$ and $\delta K_S$.

\bibitem{ashcroft76} N.W. Ashcroft and N.D. Mermin, In: {\it Solid State
Physics} (Saunders College Publishing, New York 1976).

\bibitem{remark2} More than just the doping dependence is contrary to
experiment: the slope above $T_c$ has a magnitude which is
correlated with the magnitude of the anomaly, but not a sign change,
as we have here.

\bibitem{remark3} These curves differ roughly by a factor of 2 from those presented in
Ref. (\onlinecite{carbone06b}) because of a difference in
definition.

\bibitem{kaminski06} A. Kaminski, S. Rosenkranz, H.M. Fretwell, M.R. Norman,
M. Randeria, J.C. Campuzano, J-M. Park, Z.Z. Li, and H. Raffy, Phys.
Rev. B {\bf 73}, 174511 (2006).

\bibitem{nuss91} M.C. Nuss, P.M. Mankiewich, M.L. O'Malley, E.H.
Westerwick, and P.B. Littlewood, Phys. Rev. Lett. {\bf 66}, 3305
(1991).

\bibitem{kuroda90} Y. Kuroda and C.M. Varma, Phys. Rev. B {\bf 42}, 8619 (1990).

\end{thebibliography}

\end{document}